\documentclass[journal = jctcce, manuscript = article]{achemso}
\usepackage{color, array, dcolumn}
\usepackage{multirow}
\usepackage{amsmath}
\usepackage{amssymb}
\usepackage{siunitx}
\usepackage{braket}
\usepackage{adjustbox}

\title{Improving the description of interlayer bonding in TiS$_2$
by Density Functional Theory}
\author{Matteo Ricci}
\author{Alberto Ambrosetti}
\author{Pier Luigi Silvestrelli}
\email{psil@pd.infn.it}
\affiliation{Dipartimento di Fisica e Astronomia ''G. Galilei'', Universit\`a di Padova,  via Marzolo 8, I--35131, Padova, Italy.} 

\begin{document}
\newpage
\begin{abstract}
\noindent
We investigate energetic and electronic properties of TiS$_2$, an
archetypal van der Waals (vdW) material, from first principles,
in the framework of the Density Functional Theory (DFT).
In this system a recent experimental study showed a puzzling
discrepancy between the distribution of the electron density in the
interlayer region obtained by X-ray diffraction data and that 
computed by DFT, even adopting DFT functionals that should
properly include vdW effects. Such a discrepancy
could indicate a partial failure of state-of-the-art DFT approaches
in describing the weak interlayer interactions of TiS$_2$ and, possibly,
of similar systems too. In order to shed light on this issue, we have
carried out simulations based on different DFT functionals, basically
confirming the mentioned discrepancy with the experimental findings.
Subsequently, we have tried to reproduce the experimental interlayer
electronic density deformation both by changing the parameters 
characterizing the rVV10 DFT functional (in such a way to 
artificially modify the strength of the vdW 
interactions at short or long range), and also by adopting a
modified pseudopotential for Sulfur atoms, involving $d$ orbitals. 
The latter approach turns out to be particularly promising. 
In fact, using this novel, more flexible
pseudopotential, we obtain not only an electronic density deformation
closer to the experimental profile, but also a better estimate of the
interlayer binding energy. Interestingly, this improvement in the
theoretical DFT description is not limited to TiS$_2$ but also applies
to other similar layered systems involving S atoms, such as  
TaS$_2$, HfS$_2$, and MoS$_2$.
\end{abstract}

\maketitle
\newpage                                                                             
\section{Introduction}
\noindent
First identified in 1873,\cite{vdW} the van der Waals (vdW) interactions are forces that today attract more interest than ever. They are present everywhere, but their manifestations still pose challenging questions which are relevant for such varied systems as soft matter, surfaces, and DNA, and in phenomena as different as supramolecular binding, surface reactions, and the dynamic properties of water. vdW interactions play a central role even in the characterization of layered materials, such as transition metal dichalcogenides (TMDs).
These represent a large class of materials with mild stiffness, which are not as soft as tissues and not as strong as metals.\cite{LM}
They are two-dimensional systems which have unique properties and 
are central to current research in solid-state science. 
In TMDs, a transition metal atom (M) layer is sandwiched between two chalcogen 
atom (X) layers and it is commonly assumed that the MX$_2$ slabs are stacked 
by vdW interactions, whereas the intralayer M-X interactions are instead 
covalent. The weak, interlayer vdW interactions play a key role in the 
formation, intercalation, exfoliation and layer-by-layer building of TMD 
materials, as well as being decisive for their characteristic properties.\\
\noindent
Like all non-relativistic electronic effects, the vdW interactions are present in the (unknown) exact Density Functional Theory (DFT) exchange-correlation (XC) functional, but they are not properly described by standard approximate DFT schemes, such as local density approximation (LDA)\cite{LDA} or semilocal generalized gradient approximation (GGA).\cite{PBE}
Hence the need to introduce truly non local XC functionals, able to accurately account for vdW corrections, arises.
Despite a considerable amount of work focused on the development of accurate vdW-corrected DFT methods, a precise characterization of the layer-layer distance and of the interlayer binding energy of vdW layered materials, still represents a challenging issue.\cite{Kohn,Pernal,Stohr,Ruiz,Su}
This is due to the nonlocal and long-range nature of the vdW interaction, as well as to the coexistence of the weak vdW bonding and much stronger intralayer chemical bonding. Fully accounting for the vdW interaction is achievable by high-level methods such as quantum Monte Carlo (QMC),\cite{Foulkes} coupled-cluster singles and doubles with perturbative triples (CCSD(T)),\cite{Raghavachari} and exploiting the adiabatic-connection fluctuation-dissipation theorem within the random-phase approximation (RPA),\cite{Eshuis} which are, however, only feasible for limited-size systems because of their high computational cost.\\

The physical quantity lying at the DFT core, is the electron density (ED), 
which fully determines the total energy of any many-electron system 
in its ground state.
The ED distribution probably represents the most information-rich observable available since it has become possible to determine EDs from analysis of structure factors obtained from accurate X-ray diffraction data. During the past decade, the accuracy of experimental X-ray diffraction data has increased dramatically owing to the use of high-energy synchrotron sources, which significantly limit systematic errors in the data and, thanks to these modern techniques, nowadays it is possible to evaluate both structural and electronic properties of a wide class of layered materials in an extremely accurately way.
In this respect, Medvedev \textit{et al.}\cite{Medvedev} have pointed out that modern DFT functionals are constructed on the basis of empirical fitting on energetic and geometrical benchmarks, neglecting the ED as a parameterization parameter.
If, from one side, this leads to improved predictions of energetic and structural features, from the other, the exact reproduction of the ED has worsened.
In a recent study, Kasai \textit{et al.}\cite{Kasai} studied the archetypal vdW material, TiS$_2$, to compare the ED distribution derived from X-ray diffraction data with the theoretical one obtained by several DFT functionals. They obtained a good agreement between theory and experiment for the description of the intralayer, covalent Ti-S interaction, but, at the same time, significant discrepancies were found for the interlayer vdW interactions, particularly considering the ED distribution in the region between S atoms belonging to adjacent layers. In fact, while
the properties at the bond critical point (BCP) were quite similar, noticeable differences were instead observed away from the BCP: considering the theoretical DFT density, the S atoms behave as if they are practically not interacting with the neighboring layer, showing charge concentration and accumulation in the region expected for a lone pair in an \textit{sp3}-hybridized S atom with three bonds to nearby Ti atoms and no other bonds; in the experiment, however, an appreciable charge accumulation and concentration was observed in the interlayer region, which was interpreted as a sign of a stronger and more directional interaction between S atoms. 
Thus, this study suggested the inability of current DFT functionals to accurately describe the interlayer ED for vdW layered materials, thus supporting the conclusions of Medvedev \textit{et al.} 
In order to shed light on this issue, in the first part of this study, we apply the LDA, the semilocal PBE\cite{PBE} and the vdW-corrected rVV10\cite{rVV10} DFT functionals, with standard Projector-Augmented Wave (PAW) pseudopotentials,\cite{PAW} to explore the behavior of the interlayer ED in different regimes, obtained by also modifying the parameters of the rVV10 functional, so that to artificially tuning the intensity of the vdW interactions. The basic result is that, in order to produce an ED
distribution appreciably closer to the experimental profile, we have to 
select the rVV10 parameters in such a way to unphysically increase the
strength of the vdW interactions, thus leading to an exaggeratedly high 
interlayer binding energy. 
In the second part a standard rVV10 approach is used, but a novel 
pseudopotential is adopted, which explicitly accounts for the 3$d$ orbitals of the S atom. While these orbitals are empty in the isolated, neutral S atom, they
can play a role\cite{dstate} in molecules and condensed-matter systems where S atoms interact with other atoms and form bonds. Interestingly, 
using this novel, more flexible S pseudopotential, we obtain not only an ED 
distribution closer to the experimental profile, but also a better estimate 
of the interlayer binding energy. We finally show that this improvement in 
the theoretical DFT description is not limited to TiS$_2$ but also applies
to other, similar layered systems involving S atoms, such as  
TaS$_2$ HfS$_2$, and MoS$_2$.

\section{Theoretical background and Computational details}
\noindent
In DFT the properties of a many-electron system can be determined by introducing suitable functionals of the electron density $n$(\textbf{r}); the most interesting many-body physical effects are described by the so-called exchange-correlation (XC) functional, related to the XC potential: $V_{xc}[n(\textbf{r})]  = V_{x}[n(\textbf{r})] + V_{c}[n(\textbf{r})]$.
In principle $V_{xc}$ describes all the many-body effects, including the
vdW interactions, however, in practice, a number of approximation must be introduced, since, unfortunately, the exact form of the XC functional is not known. Nowadays several kinds of approaches to approximately include vdW interaction in DFT calculations are available. Some of these are based on semiempirical, atom-based pair potentials, able to give approximate vdW corrections;\cite{grimme1,grimme2,xdm} instead others are built introducing a suitable non-local XC density functional, such as the vdW-DF family by Dion \textit{et al.}\cite{Dion} and the scheme proposed by Vydrov and Van Voorhis,\cite{VV10} with the revised, more efficient version rVV10.\cite{rVV10} 
rVV10 certainly represents one of the best vdW-corrected functionals nowadays available; it takes into account the entire range of vdW interactions at a reasonable computational cost using only the ED $n$(\textbf{r}) as input.
In particular, the rVV10 non-local correlation energy has the following expression:
\begin{equation}
E_c^{NL}\,=\, 
\int\int\,n(\textbf{r})\Phi(\textbf{r}\,,\textbf{r'}) n(\textbf{r'})\label{1}
\end{equation}
where $\Phi$ is the non-local correlation kernel, which is:
\begin{equation}
\Phi = -\frac{3e^4}{2m^2} \frac{ 1 } 
{(q(\textbf{r}) R^2 + 1)(q(\textbf{r'}) R^2 + 1)
[(q(\textbf{r})\,+\,q(\textbf{r'})) R^2 + 2]}			\label{2}
\end{equation}
In the above expression $R = | \textbf{r} - \textbf{r'} |$ and $q(\textbf{r})$ is a function of the ED and its gradient:
\begin{equation}
q(\textbf{r}) =  
\frac{\omega_0 (n(\textbf{r}), \nabla n(\textbf{r}))}
{k(n(\textbf{r}))}
\end{equation}
where 
\begin{equation}
\omega_0 = 
\sqrt{ C \frac{\hbar^2}{m^2}  
\left|\frac{ \nabla n(\textbf{r})} {n(\textbf{r})}\right|^4 + 
\frac{1}{3} \frac{4\pi n(\textbf{r}) e^2}{m}}
\label{C}
\end{equation}
and
\begin{equation}
k(n (\textbf{r})) = 3 \pi b \left(\frac{n(\textbf{r})}{9 \pi} \right)^{1/6} \label{B}\;.
\end{equation}
The total XC functional is then obtained by adding the non-local correlation energy to the refitted Perdew-Wang exchange functional\cite{rPW86} and the Perdew-Wang LDA correlation\cite{LDAc} functional, as proposed in the seminal work by Vydrov and Van Voorhis:\cite{VV10}
\begin{equation}
E_{xc}^{NL}\,=\,E_{x}^{rPW86} + E_{c}^{LDA} + E_{c}^{NL}\;.
\end{equation}
The rPW86 exchange functional was chosen mainly because it is nearly vdW-free,\cite{rPW86} in such a way to avoid double-counting effects. In eq. \eqref{C} the $C$ parameter represents the so-called local band gap and is tuned to give accurate asymptotic vdW $C_6$ coefficients, thus regulating the behavior of the long-range component of vdW interactions. The fitting procedure, aimed to minimize the average error for a benchmark set of 54 $C_6$ coefficients, was originally carried out by Vydrov and Van Voorhis,\cite{VV10} leading to an optimal value of $C$ = 0.0093.
Another essential aspect of the VV10 (and its successor rVV10) formalism is the presence of a second adjustable parameter, $b$ (see eq. \eqref{B}), which controls the short range behavior of the non-local correlation energy. 
This means that when E$_c^{NL}$ is added to the other energetic terms, the $b$ parameter is adjusted to merge the interaction energy contributions at short and intermediate ranges.
With an empirical fitting procedure on the S22 binding-energy data set,\cite{BE} Vydrov and Van Voorhis proposed a value of $b$ = 5.9.\cite{VV10} After the implementation of the efficient Roman-Perez, Soler\cite{RPS} interpolation scheme in the reciprocal space, this value was revised by Sabatini and coworkers to $b$ = 6.3.\cite{rVV10}\\
We have chosen rVV10 as the basic DFT functional to investigate the effect of vdW interactions in the ED distribution of TiS$_2$ (although extensive tests were
also performed using other vdW-corrected functionals), also because of the
possibility to separately modify the intensity of both short- and long-range vdW interactions in a transparent way by simply tuning the two adjustable parameters $b$ and $C$ mentioned above.
First the short-range component is analyzed. This is done by considering two different regimes: the former in which the value of the $b$ parameter is increased to $b$=10.0, thus damping the intensity of short range vdW interactions, and the latter in which $b$ is reduced to $b$=1.0, that instead results in a substantial increase of the intensity of the short-range vdW interaction. 
While tuning the parameters of the rVV10 functional can improve the
description of some properties of the TiS$_2$ system, clearly this could
lead to a reduction of the transferability of the modified functional
to other systems.

Our ab-initio calculations have been performed using the version 6.5 of the Quantum ESPRESSO (QE) DFT package.\cite{QE1,QE2}
We first used for Ti and S atoms standard PAW\cite{PAW} pseudopotentials taken 
from the QE database.\cite{pseudo}
Subsequently, for further testing, we have generated, for the S atom, 
a novel pseudopotential, using the \textit{atomic} QE program.
We report DFT results obtained using the LDA, the semilocal GGA PBE 
functional, and the vdW-corrected functional rVV10, 
both in its standard form and considering the
variants (parameter changes and combination with a novel S pseudopotential)
described above.
Self-consistent calculations and relaxation processes, leading to the ground-state equilibrium geometry of bulk TiS$_2$, were carried out within the \textit{PWscf} QE program. The calculations adopted plane-wave and density cutoffs of 100 and 1000 Ry, respectively, to get highly-converged results, a cold-smearing parameter of 0.01 Ry, and a 12$\times$12$\times$4 k-point mesh for the sampling of the Brillouin Zone (BZ). 
The ED analysis is carried out using the \textit{PostProc} QE program, employing the B-Spline method\cite{spline} for the $n(x, y, z)$ interpolation; in particular, for the ED profile computed along the S-S, S-center, and Ti-S lines, a real-space grid of 600$\times$600$\times$800 (x, y, z) points, generated inside a parallelepiped containing the axis of interest, is used.
To get a reliable estimate of the $\nabla^{2}n$(\textbf{r}) function, in the case of the S-S line, the distance between the parallelepiped axis and the edges of its sides is 0.5 \si{\angstrom}, for the S-center case 0.25 \si{\angstrom}, and finally 0.35 \si{\angstrom} for the Ti-S line.
The ED is then processed by using the \textit{Critic2} program.\cite{Critic2_1,Critic2_2} 
More specifically, we compute the ED charge deformation, 
$\Delta n$(\textbf{r}), by considering a spatial mesh of 72$\times$72$\times$108 points inside the unit cell (containing one Ti atom and two S atoms) 
and taking the difference between the ED $n$(\textbf{r}) and the
sum of the EDs of the isolated Ti and S atoms.   
$\Delta n$(\textbf{r}) profiles and Laplacian $\nabla^{2}n$(\textbf{r}) maps, in the interlayer and intralayer areas, are then evaluated on the plane of interest using \textit{Critic2}. The calculation of the atomic properties of a given S atom is carried out by making an integration of the so-called ''atomic basin'' of the S atom, represented by the space region whose surface has zero-density gradient flux.
This is done again with the \textit{Critic2} program, that allows to compute spherical multipolar moments defined as
\begin{equation}
Q_{lm} = \int_A n(\textbf{r})\,r^l\,Y_{lm}
\label{m1}
\end{equation} 
where the integration domain $A$ is the atomic basin volume. In the above equation, $n(\textbf{r})$ is the charge ED and $Y_{lm}$ are the spherical harmonics, in which the $m$ index runs over the interval [-$l$\,;\,$l$]. With this definition, the magnitude of the dipole moment attributed to the S atom is obtained as the square root of the sum of the squared $l$=1 terms:
\begin{equation}
d(q_A) = \sqrt{\sum_{m=-1,0,1} Q_{1m}^2}\,\;. 
\end{equation}

\section{Results and discussions}
\noindent
Starting from the experimental lattice parameters, $a$ = 3.398 \si{\angstrom} and $c$ = 5.665 \si{\angstrom}, as reported by Kasai \textit{et al.},\cite{Kasai} we relax the crystal structure in the intralayer atomic plane until the force acting on the system is smaller than $10^{-5}$ Ry/a.u. This relaxation process, carried out with both PBE and  rVV10 functionals, leads to very similar results, 
as expected,\cite{Kasai} since vdW interactions do not substantially influence the geometry of the intralayer plane. 
By further optimizing the interlayer geometry, the standard rVV10 functional
predicts that
the distance between two S atoms belonging to adjacent layers is 3.443 \si{\angstrom} while the intralayer S-Ti distance is 2.421 \si{\angstrom}, in excellent agreement with the reference DFT values (at the SCAN+rVV10 level\cite{SCAN+rVV10}) of 3.446 \si{\angstrom} and 2.420 \si{\angstrom}, respectively, reported in ref. \citenum{Kasai}.
\begin{figure}[!h]
\begin{center}
\includegraphics[width=\linewidth]{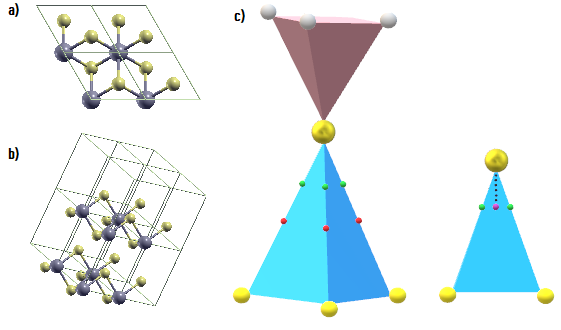}
\end{center}
\caption{\scriptsize a) \& b): bulk structure of TiS$_2$. The unit cell is repeated twice along the $x$, $y$, and $z$ directions. c) On the left hand side the TiS$_2$ octahedron is represented: the bigger S atom in the middle is the reference S atom, while the grey balls denote Ti atoms. The lower tetrahedron sides indicate the S-S lines, connecting the reference atom to the S atoms belonging to the adjacent layer. On the right hand side, the S-center line (connecting the reference S atom to the center of the lower tetrahedron) is represented by dots. The green spheres denote the positions of the (3, -3) critical points (CPs) along the S-S lines, the red spheres are referred to the (3, -1) CPs and, finally, the violet sphere shown inside the triangular section (right side of the c) panel) represents the position of the (3, -3) CP belonging to the S-center line.}
\end{figure}
As already mentioned, Kasai \textit{et al.}\cite{Kasai} pointed out that
DFT simulations are characterized by a 
an interlayer S-S ED deformation significantly lower than the experimental one, implying that the X-ray diffraction data predict a stronger S-S interlayer interaction than DFT.
Moreover, the experimental ED is characterized by 3 maxima in the direction of the 3 neighboring, symmetry-related S atoms, while the theoretical one exhibits only one maximum directed towards the center of the S-atom tetrahedron (see Figure 1), suggesting that the 
theoretical DFT description of the interlayer interactions indeed 
qualitatively differs from the experimental one.
In order to investigate the reasons of this discrepancy, we analyze the results of our DFT calculations and focus on the two directions of main interest, namely the S-center and the S-S line. Using the Bader nomenclature, for each line we characterize the (3, -3) critical point (CP) evaluating the position, \textbf{r}$_{max}$, where the charge density reaches its maximum value, the density at this point, 
$n$(\textbf{r}$_{max}$), and the Laplacian, $\nabla^{2}n$(\textbf{r}$_{max}$).
Furthermore, we do the same for the (3, -1) CP, computing \textbf{r}$_{min}$, 
$n$(\textbf{r}$_{min}$), and $\nabla^{2}n$(\textbf{r}$_{min}$). 
As a preliminary analysis, we have computed the ED and the Laplacian along the Ti-S line, identifying the main features of the polar covalent bond in the intralayer plane. 
The values that characterize the ED are reported in Table S1 of 
\textit{Supporting information}.
Two non-equivalent (3, -3) CPs are found along the Ti-S line (see top left panel in Figure S1 of \textit{Supporting information}). 
The (3, -1) CP is instead located at 1.12 \si{\angstrom} from the Ti atom. 
As can be seen, looking at Table S1 of \textit{Supporting information}, very similar values are predicted by 
all the DFT functionals applied and most of these are also in good agreement with the experimental data.\\
A similar analysis can be also applied to characterize the interlayer bonding:
the properties of the (3, -3) CP for both the S-S and S-center lines, and for the (3, -1) CP along the the S-S line are summarized in Table 1 and 2, respectively. 
In the interesting case of the S-S line, the PBE, rVV10, and rVV10($b$=10.0) functionals predict a maximum value of the ED at a distance \textbf{r}$_{max}$ $\sim$ 0.61 \si{\angstrom} from the reference S atom, with a relative error of $\sim$ 13 \% with respect to the experimental result,\cite{Kasai} while the error is slightly larger with rVV10($b$=1.0) and LDA functionals. 
\begin{table}[h!]
\caption{\scriptsize Properties at the (3, -3) CP along the S-S and S-center lines, evaluated using different DFT functionals.}
\begin{center}
\begin{tabular}{l| l| c| c| c}
&line&\textbf{r}$_{max}$ (\si{\angstrom}) & $n$(\textbf{r}$_{max}$) (e/\si{\angstrom}$^3$) & 
$\nabla^{2}n$(\textbf{r}$_{max}$) (e/\si{\angstrom}$^5$) \\
\hline
\multirow{2}{*}{rVV10}&S - S&0.614&1.102&-9.163\\
					  &S - center&0.614&1.132&-10.052\\
\hline
\multirow{2}{*}{rVV10($b$=1.0)}&S - S& 0.607& 1.050&-8.663\\
							  &S - center&0.605&1.081&-9.479\\
\hline
\multirow{2}{*}{rVV10($b$=10.0)}&S - S&0.614&1.105&-9.222\\
							   &S - center&0.614&1.132&-9.985\\
\hline
\multirow{2}{*}{PBE}&S - S&0.614&1.095&-9.093\\
					&S - center&0.614&1.129&-9.934\\
\hline
\multirow{2}{*}{LDA}&S - S&0.579&1.141&-10.016\\
					&S - center&0.579&1.172&-11.419\\
\hline
\multirow{2}{*}{rVV10$opt$}&S - S&0.609&1.074&-8.950\\
						 &S - center&0.607&1.102&-9.766\\
\hline
Experiment\cite{Kasai}&S - S&0.707&1.150&-9.460\\
Theory\cite{Kasai}(SCAN+rVV10)&S - center&0.711&1.080&-8.910\\
\hline
\end{tabular}
\end{center}
\end{table}

\begin{figure}[!h]
\begin{center}
\includegraphics[width=\linewidth]{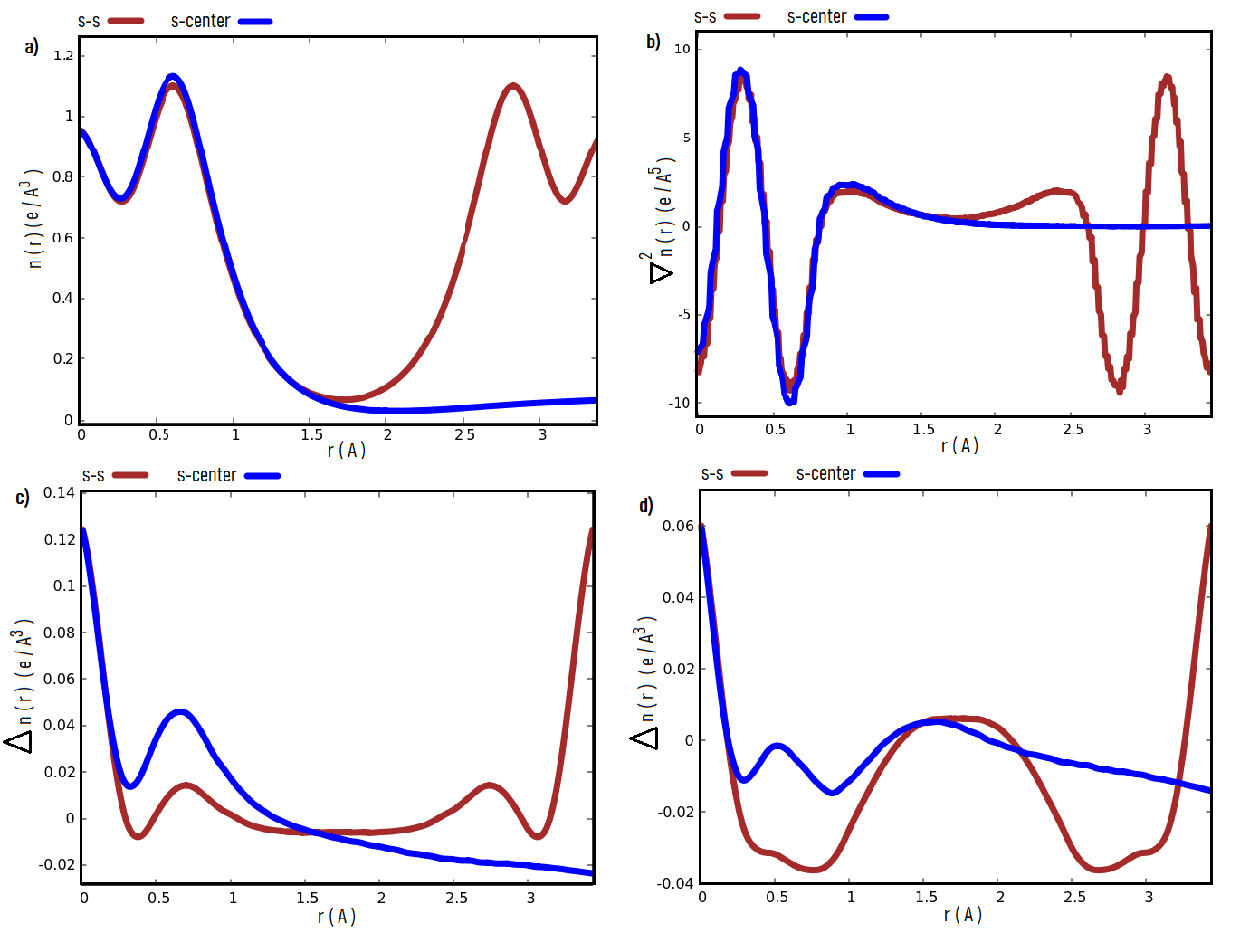}
\end{center}
\caption{\scriptsize a) ED computed along the S-S and 
S-center line with the PBE functional. b) Corresponding Laplacian profiles. The small oscillations of the Laplacian profile are due to the finite-size of the real space mesh considered for the ED computation.
In this case $\nabla^{2}n$(\textbf{r}$_{max}$) and $\nabla^{2}n$(\textbf{r}$_{min}$) are estimated by a parabolic interpolation $f(r \sim c) = c_2 (r-c)^2 + k$, where 
the expansion center is $c$ =  \textbf{r}$_{max}$ or $c$ = \textbf{r}$_{min}$.
c) ED deformation $\Delta$ $n$(\textbf{r}) computed using the PBE functional and d) the rVV10($b$=1.0) functional. Curves obtained by rVV10 and rVV10($b$=10.0) functionals are very similar to the PBE ones and therefore are not reported.}
\end{figure}
\newpage
The $n$(\textbf{r$_{max}$}) values are all very similar and in reasonable agreement with the experimental data, with the LDA functional that shows the best agreement with the experimental value. The rVV10($b$=1.0) result, for which the maximum discrepancy occurs, has a deviation of only $\sim$ 9 \%.
The Laplacians $\nabla^{2}n$(\textbf{r}$_{max}$) are also in acceptable agreement with the experimental values. Again, the result of the rVV10($b$=1.0) functional exhibits the maximum deviation from the experiment, with a discrepancy of $\sim$ 8 \%.
This means that, at a distance r$_{max}$ from the reference S atom, along the S-S line, with rVV10($b$=1.0), the ED deformation is slightly smaller than for the other functionals.\\

For the S-center line a direct comparison with experimental data is not available, so we take as reference the theoretical SCAN+rVV10 result reported by Kasai \textit{et al.}\cite{Kasai}
The S-center line is characterized by the same \textbf{r}$_{max}$ value found for the S-S line, again with a small deviation of the value obtained with rVV10($b$=1.0) and LDA. Even in this case, the results for the ED and the Laplacians are in line with the theoretical reference data.\cite{Kasai} Interestingly, for each functional employed, a slightly larger 
$n$(\textbf{r$_{max}$}) value than that relative to the S-S line is observed, with a stronger ED deformation described by the associated Laplacian, indicating that the charge deformation along the S-center line is stronger than for the S-S line.
The maximum deviation from the reference values, for both ED and Laplacians, occurs again for the LDA functional with an error of $\sim$ 8 \% and $\sim$ 28\%, respectively.\\

\begin{table}[h!]
\caption{\scriptsize Properties at the (3, -1) CP along the S - S line, evaluated using different DFT functionals. The theoretical (SCAN+rVV10) result reported, is the result achieved after a multipolar refinement of the pure DFT density (see ref. \citenum{Kasai} for details).}
\begin{center}
\begin{tabular}{l| c| c| c}
&\textbf{r}$_{min}$ (\si{\angstrom}) & $n$(\textbf{r}$_{min}$) (e/\si{\angstrom}$^3$) & 
$\nabla^{2}n$(\textbf{r}$_{min}$) (e/\si{\angstrom}$^5$) \\
\hline
rVV10&1.718&0.068&0.486\\
rVV10($b$=1.0)&1.718&0.079&0.461\\
rVV10($b$=10.0)&1.718&0.067&0.481\\
PBE&1.718&0.067&0.460\\
LDA&1.718&0.070&0.468\\
rVV10$opt$&1.718&0.074&0.464\\
\hline
Theory\cite{Kasai}(SCAN+rVV10)& &0.058&0.727\\
Experiment\cite{Kasai}& &0.086&0.691\\
\hline
\end{tabular}
\end{center}
\end{table}
\newpage
\begin{figure}
\begin{center}
\includegraphics[width=\linewidth]{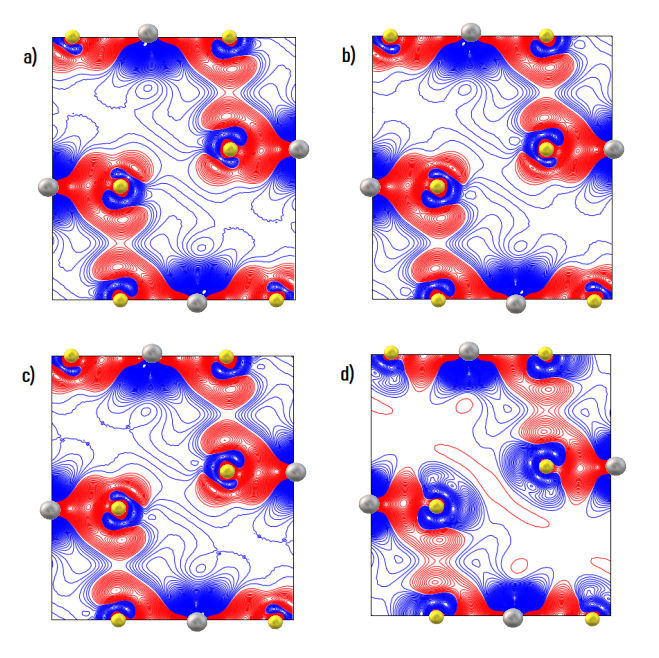}
\end{center}
\caption{\scriptsize ED deformation map in the interlayer plane using four different functionals: a) PBE, b) rVV10, c) rVV10($b$=10.0), and d) rVV10($b$=1.0). The Ti atom are represented by grey spheres, while the S atoms are the yellow spheres. In the present maps, the smallest isoline value reported is 0.001 e/\si{\angstrom}$^3$, with red and blue contours that represent charge accumulation and depletion, respectively.}
\end{figure}

The S-S bond shows a symmetric ED distribution along the line connecting the two S atoms (see Figure 2a)), with \textbf{r}$_{min}$ exactly halfway between the two S atoms. 
Our computed $n$(\textbf{r}$_{min}$) values, with the PBE, rVV10, and rVV10($b$=10.0) functionals, as well as LDA, are in very good agreement with the DFT densities presented in ref. \citenum{Kasai}, exhibiting a significant underestimate (by about 22 \%) of the ED at the interlayer BCP, with respect to the experimental value.
Interestingly, using the rVV10($b$=1.0) functional, the discrepancy with the
experiment is instead considerably reduced to about 8 \%.
Note that the reported, theoretical SCAN+rVV10 value, \cite{Kasai} is slightly below (and therefore slightly worse than) all our results.
For the (3, -1) CP, the Laplacian sign is always positive, as expected, indicating that the interlayer BCP located at \textbf{r}$_{min}$ is a point where a charge depletion occurs. 
In Figure 2a) and 2b), an example of ED profiles and Laplacians, at the PBE level, along the S-S and S-center lines are shown, while in the bottom part, a comparison between the PBE ED deformation $\Delta n$(\textbf{r}) (Figure 2c)) and that obtained by rVV10($b$=1.0) (Figure 2d)) is made along the same lines, highlighting that rVV10($b$=1.0) predicts a stronger charge depletion in the vicinity of the reference S atom than PBE and the other functionals. A consequent charge accumulation in the interlayer region appears, as confirmed by the higher ED $n$(\textbf{r}$_{min}$)$_{\mbox{S}\,-\,\mbox{S}}$ at the (3, -1) CP accompanied by a smaller Laplacian value $\nabla^2 n$(\textbf{r}$_{min}$)$_{\mbox{S}\,-\,\mbox{S}}$ (see Table 2) that indicates a slightly less charge depletion than for the other functionals considered. 

The ED deformation map in the interlayer plane, presented in Figure 3, is very similar with the PBE, rVV10, and rVV10($b$=10.0) functionals (as well as the LDA case which it is not reported), while the same is not true for rVV10($b$=1.0), which predicts a more pronounced ED deformation in the region between the two S atoms belonging to adjacent layers. 
The Laplacian map of the interlayer plane is presented in Figure S2 of the \textit{Supporting information}: in this case, no significant differences for the functional schemes adopted up to now are found, as well as for LDA and the other schemes presented below. 

\begin{table}[h!]
\caption{\scriptsize Cohesive energy (CE) and interlayer binding energy (ILBE), see text for the definitions, obtained using the experimental lattice parameter $c$ (first and second column) and the  
corresponding quantities (denoted by the $^{*}$ symbol, in the fourth and
fifth column) relative to the $c$ value (third column) optimized for each adopted functional. The reported reference CE and $c$ values are experimental data, while, for the ILBE, we provide theoretical RPA\cite{bj} and MBD estimates.
The quantity V, in the sixth column, represents the total charge belonging to the atomic basin $A$, surrounding the S atom. The charge q$_{A}$, contained in $A$, and the atomic dipole moment of the S atom, d(q$_{A}$), are reported in column seven and eight respectively.}
\begin{center}
\begin{adjustbox}{max width=\textwidth}
\begin{tabular}{l| c| c| c| c| c| c| c| c}
&CE (\si{\electronvolt})&ILBE (m\si{\electronvolt}/\si{\angstrom}$^2$)
&$c$ (\si{\angstrom})&CE$^{*}$ (\si{\electronvolt})&ILBE$^{*}$ (m\si{\electronvolt}/\si{\angstrom}$^2$)&V (\si{\angstrom}$^3$)&q$_{A}$ (e)&d(q$_{A}$) (e\si{\angstrom})\\
\hline
rVV10&-15.602&-27.828&5.728&-15.603&-27.888&23.775&-0.860&0.380\\
rVV10($b$=1.0)&-27.928&-192.073&4.934&-42.140&-1613.663&23.781&-0.861&0.383\\
rVV10($b$=10.0)&-15.035&-16.609&5.878&-15.045&-17.698&23.574&-0.796&0.324\\
PBE&-15.150&-3.013&6.595&-15.189&-0.861&23.804&-0.871&0.404\\
LDA&-18.633&-17.275&5.470&-18.633&-18.357&23.648&0.829&0.383\\
rVV10$opt$&-25.889&-151.580&5.123&-25.310&-175.506&23.615&-0.809&0.373\\
\hline
Experiment&-14.805\cite{ce}& &5.665\cite{Kasai}&-14.805& &23.470\cite{Kasai}&-0.820\cite{Kasai}&0.030\cite{Kasai}\\
Theory (RPA)& &-18.900\cite{bj}& & &-18.900&23.630\cite{Kasai}&-0.800\cite{Kasai}&0.340\cite{Kasai}\\
Theory (MBD)& &-19.031& & &-19.031& & & \\
\hline
\end{tabular}
\end{adjustbox}
\end{center}
\end{table}
\newpage
A summary of the basic energetic quantities of TiS$_2$ and the atomic 
properties of the reference S atom is reported in Table 3.
We optimized the interlayer lattice constant $c$ for each functional adopted (third column of Table 3), obtaining results close to the experimental reference value using the rVV10 and rVV10($b$=10.0) functionals, with an error of only $\sim$ 3 \%. Instead, the quality of the PBE, rVV10($b$=1.0), and LDA estimates is worse, with PBE that substantially overestimates, and LDA and rVV10($b$=10.0) which instead underestimate. This behavior does not come as a surprise: in fact LDA and the semilocal PBE functional are not able to properly describe the vdW interactions, while the rVV10($b$=1.0) functional was deliberately built to artificially increase the strength of short-range vdW interactions, thus leading to a larger interlayer bonding energy and to a shorter layer-layer distance.
The atomic cohesion energy (CE) is defined as E(TiS$_2$) - E(Ti) - 2E(S), where
E(TiS$_2$) is the total energy of the TiS$_2$ systems, while E(Ti) and E(S)
are the total energies of the isolated T and S atoms. This quantity mainly
reflects the intensity of the strong, intralayer Ti-S covalent bonds.
Considering both the experimental configuration (first column) and the one in which the lattice constant $c$ is optimized (fifth column), quite similar results are achieved using
PBE, rVV10, and rVV10($b$=10.0), with an average discrepancy with the 
experimental reference value of $\sim$5 \%. 
However, the CE value obtained with the rVV10($b$=1.0) functional is largely
overestimated implying that artificially increasing the strength of the
short-range vdW interaction leads to a severe overbinding even of the 
Ti-S covalent bond. Therefore, the improvement in the ED distribution using
rVV10($b$=1.0) comes at the expense of an unacceptable deterioration in the
description of the energetic properties. 
This is also confirmed by the analysis of the interlayer binding energy (ILBE), defined as E(TiS$_2$) - E(TiS$_2$)$_{\mbox{c/a=5}}$, representing the interaction energy between two 
adjacent layers of the material. Here, E(TiS$_2$)$_{\mbox{c/a=5}}$ indicates
the total energy of the system when the $c$ parameter is set to 
$5a \sim$17 \si{\angstrom}, which corresponds to a distance so large that
the interlayer interaction is negligible.
In order to have a more sound comparison with reference values,
in addition to the RPA value reported in the literature,\cite{bj}
we have also computed the ILBE using the Many-Body Dispersion (MBD)
approach\cite{mbd} which allows for an effective RPA description of
long-ranged vdW interactions, beyond conventional pairwise approximations.
Within MBD one maps the atomic response functions into a set of
quantum harmonic oscillators, mutually coupled at the dipole-dipole level.
Due to the inclusion of many-body effects, MBD can
reach chemical accuracy for a broad variety of systems (from small molecules to large nanostructure and periodic materials)\cite{mbd}.
In particular, many-body effects were shown to be crucial in two-dimensional vdW
materials~\cite{prb}, as they can
qualitatively alter the power law scaling of dispersion interactions.
Remarkably (see Table 3) the MBD estimate is very close to the
RPA value reported in ref.\citenum{bj}.

As expected, PBE dramatically underestimates the ILBE (the discrepancy is
even worse considering the optimal configuration predicted by PBE), a
behavior that can be clearly ascribed to the inadequate inclusion of
vdW effects by PBE. As can be seen in Table 3, the standard rVV10 
functional significantly overestimates the ILBE, while rVV10($b$=10.0) gives
a result reasonably close to the reference values. Interestingly, the LDA estimation is very good, showing the best agreement with the reference values. 
However, one must point out that this good performance is accidental:
the well-known LDA overbinding due to the overestimate of
the long-range part of the exchange contribution, somehow mimics the
missing vdW interactions, thus leading to reasonable estimates of the
binding energies in a wide range of systems.\cite{LDAprob}
Instead, similarly to what observed for the CE, rVV10($b$=1.0) predicts a very large, unphysical ILBE. 

Looking at Table 3 our computed atomic basin volume V turns out to be 
in good agreement 
with the theoretical and experimental reference values, and the same is true 
for the total charge q$_{A}$ contained in the atomic basin, with all the 
adopted functionals that predict quite similar values (which the partial 
exception
of rVV10($b$=10.0) which gives a significantly less negative accumulated 
charge). The negative value of q$_{A}$ is consistent with the higher 
electronegativity of the S atom, which therefore behaves as the 
electron-acceptor in the polar covalent Ti-S bond.  
The values of the dipole moment of the S atom, d(q$_{A}$),
computed through the spherical multipolar 
expansion (see section II), are, for all the considered functionals, in 
line with the theoretical reference data but much larger than the
experimental estimate, thus reproducing the discrepancy between theory and
experiment discussed above. In more detail, PBE exhibits the
largest discrepancy, thus confirming that this functional is not adequate,
while the rVV10($b$=10.0) functional gives a slightly less pronounced dipole 
moment, suggesting a description a little closer to the experiment, 
although even the rVV10($b$=10.0) value of d(q$_{A}$) is still quite far from the experimental one.

So far, our results for the ED distribution, and for the energetic and atomic 
properties of TiS$_2$, are in line with the conclusions of 
Kasai \textit{et al.}\cite{Kasai} concerning the theoretical DFT description. 
In particular, our analysis of CPs (see Table 1), confirms that the ED predicted at DFT level along the S-S line is smaller than the one along the S-center line, at variance with the experimental evidence.
Small, quantitative differences between our calculations and the theoretical 
SCAN+rVV10 results reported by Kasai \textit{et al.}\cite{Kasai} can be easily 
explained by the different DFT functional adopted and by other 
technical differences: for example, in ref. \citenum{Kasai} a finer mesh 
of 22$\times$22$\times$13 k-points was used for the sampling of the BZ,
this choice being motivated by the need of describing core electron 
oscillations to carry out a multipole refinement procedure.\cite{Kasai}

Coming to a more detailed analysis, we observe (see Figure S2 and Figure S3 of \textit{Supporting information}) that the maps of $\Delta n$(\textbf{r}) and $\nabla^2 n$(\textbf{r}) obtained by 
rVV10($b$=10.0) are very similar to those given by the standard rVV10 
functional and also by PBE, thus suggesting that the interlayer ED 
distribution is weakly affected by an augmented damping applied to 
the short-range vdW interactions; this is also quantitatively confirmed 
by the the values of $n$(\textbf{r}) and $\nabla^2 n$(\textbf{r}$_{max}$) 
evaluated at the (3, -3) CP (see Table 1 ). A small difference can be
noticed only at the (3, -1) CP, where the rVV10($b$=10.0) value of 
$\nabla^2 n$(\textbf{r}$_{max}$) almost coincides with that obtained by
the standard rVV10 result but weakly differs from the PBE one.   
The situation is different considering rVV10($b$=1.0), namely
a functional where the short-range vdW interaction is artificially 
increased; in fact, although the ED and the Laplacian at the CPs defined
above are similar to the same quantity evaluated by the other functionals,
more pronounced differences can be observed away from the BCP, reflecting
an accumulation of the ED in the interlayer region. As a consequence, 
rVV10($b$=1.0) predicts an ILBE value which appears to be dramatically
overestimated.  
Moreover, with rVV10($b$=1.0) one can observe (Figure 3d)) a considerable ED 
deformation between the reference S atom and the other S atom belonging to 
the same layer, thus indicating a strong intralayer S-S interaction,
a feature not observed with the other functionals and absent in 
the experimental description. 
 
To overcome (at least partially) these evident shortcomings of rVV10($b$=1.0),
we have also tested another DFT functional, rVV10($b$=1.1,$C$=0.00207), named as rVV10$opt$, which is characterized by a new $C$ parameter, thus modifying the value of the 
effective C$_6$ coefficient responsible for the long-range behavior of
the vdW interactions, and also by a slightly changed $b$ parameter. 
\begin{figure}
\begin{center}
\includegraphics[width=\linewidth]{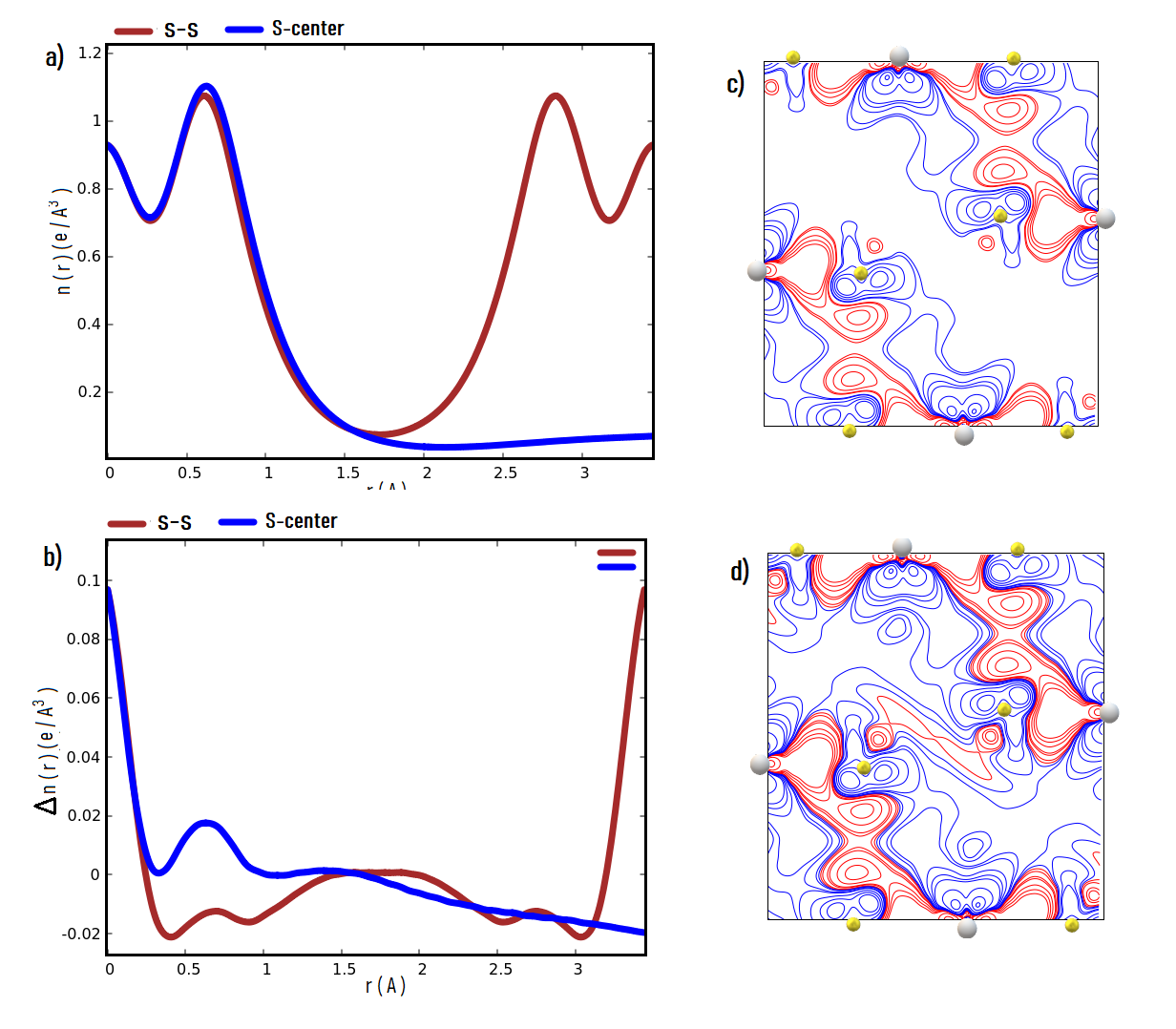}
\end{center}
\caption{ \scriptsize a) ED along the S-S and S-center line, obtained using the
rVV10$opt$ functional. b) Associated ED deformation 
$\Delta n$(\textbf{r}). c) ED deformation map in the Ti-S intralayer plane. 
d) ED deformation map in the interlayer plane with a contour of $\pm$ 0.01 e/\si{\angstrom}$^3$. e) ED deformation map in the interlayer plane with a smaller contour of $\pm$ 0.001 e/\si{\angstrom}$^3$.}
\end{figure}
The ED deformation map obtained with this new functional is shown in Figure 4. 
As can be seen, by artificially manipulating the long-range component of vdW interactions, the unphysical ED accumulation in the intralayer S-S region
disappears and, more interestingly, a small tendency to form a localized
S-S interlayer bond can be observed, although the intensity of 
the interlayer ED deformation is again underestimated, being 7 times smaller than the 
experimental one.\cite{Kasai} 
rVV10$opt$ preserves the nature of the S-Ti intralayer 
bond and the basic properties of the ED at the two non equivalent (3, -3) 
CPs and at the (3, -1) CP; in fact the positions of these points are 
unchanged and the ED and Laplacian values, reported in Table 1 are in line with the other functionals employed.
Similar conclusions hold considering the S-S and S-center lines.
Finally, one must point out that, even with this new functional, the 
ED accumulation along the S-S line is predicted to be smaller than along
the S-center line, differently from the experimental findings. 
Therefore, considering that (see Table 3 ) rVV10$opt$ still
largely overestimate the ILBE, although it slightly reduces the dramatic
overbinding of rVV10($b$=1.0), one can conclude that it is not possible
to achieve, at the same time, a description of energetic and 
ED deformation features, in reasonable
agreement with the experimental results, by simply tuning the values of the $b$ and $C$ parameters of the rVV10 functional, which determine
the strength of short/medium- and long-range vdW effects.
 
As an alternative strategy to reduce the discrepancy between the theoretical
DFT description of TiS$_2$ and the experimental evidence, we have also
investigated the effect of introducing a modified pseudopotential
for S. The S atom is characterized by the [Ne]3s$^2$3p$4$ ground-state 
configuration, so that in neutral S atom $d$ orbitals are empty, and,
in simple molecules, they are nearly unpopulated. Nonetheless, it
was shown\cite{dstate} that they can play a significant role: for instance,
even in the small S$_2$ molecule an appreciable improvement in the
binding-energy estimate was observed by adopting for S a pseudopotential
built by taking into account $d$ orbitals, which are expected\cite{dstate} to
give S enhanced polarizability and enable a great variety of bonding 
configurations through hybridization. 
In fact, although some orbitals (for instance, in this case the $d$ ones)
are not bound states of the neutral atom, it is necessary\cite{dstate}
to choose a partially ionized atomic reference configuration, with a
charge slightly different from the neutral atom, to get
pseudopotentials that are better tailored for applications to
solid-state and extended systems, since the effective configuration
of these systems differs from that of the neutral atom.

Following this approach, 
we have generated  a novel S ultra-soft pseudopotential, 
based on the configuration [Ne]3s$^2$3p$^{3.0}$3d$^{1.0}$, using the 
Troullier-Martins pseudization method\cite{tm} with non-linear core corrections.\cite{nlcc} Coupling it with the rVV10 functional, we have developed the rVV10$d$ scheme.

First we have applied the rVV10$d$ functional to the simple case of the S$_2$ molecule, which is in itself an interesting system, being an ubiquitous intermediate in the combustion, atmosphere, and interstellar space.\cite{s2} 
With rVV10$d$, while the equilibrium distance is unchanged with respect to 
rVV10, we obtain a substantial improvement of the molecular 
binding-energy and a slight improvement of the vibrational frequency
(see table S3 of \textit{Supporting information}),
thus confirming the importance of including the $d$ orbitals in such a system. 
In particular, considering the binding energy, the more than 20 \% discrepancy with the experimental reference value of the rVV10 original functional, reduces to only about 3 \% with rVV10$d$. We have therefore recomputed the energetic and electronic properties of TiS$_2$ using the new rVV10$d$ functional.

\begin{table}[h!]
\caption{\scriptsize Properties at the (3, -3) CPs along both the S-S and S-center line and at the (3, -1) CP on the S-S line, computed using rVV10$d$ and compared with the results achieved with rVV10.}
\begin{center}
\begin{tabular}{ l| l| c| c| c }
(3, -3) CPs & & & &\\
&line&\textbf{r}$_{max}$ (\si{\angstrom}) & $n$(\textbf{r}$_{max}$) (e/\si{\angstrom}$^3$)& 
$\nabla^{2}n$(\textbf{r}$_{max}$) (e/\si{\angstrom}$^5$) \\
\hline
\hline
\multirow{2}{*}{rVV10}&S - S&0.614&1.102&-9.163\\
					  &S - center&0.614&1.132&-10.052\\
\hline
\multirow{2}{*}{rVV10$d$}&S - S&0.643&1.094&-9.253\\
					     &S - center&0.643&1.126&-9.990\\
\hline
Experiment\cite{Kasai}&S - S&0.707&1.150&-9.460\\
Theory\cite{Kasai}(SCAN+rVV10)&S - center&0.711&1.080&-8.910\\
\hline
\hline
(3, -1) CP & & & &\\
&line&\textbf{r}$_{min}$ (\si{\angstrom}) & $n$(\textbf{r}$_{min}$) (e/\si{\angstrom}$^3$)& 
$\nabla^{2}n$(\textbf{r}$_{min}$) (e/\si{\angstrom}$^5$) \\
\hline
\hline
rVV10&S - S&1.718&0.068&0.486\\
rVV10$d$&S - S&1.718&0.068&0.482\\
\hline
Experiment\cite{Kasai}&S - S& &0.086&0.691\\
Theory\cite{Kasai}(SCAN+rVV10)&S - S& &0.058&0.727\\
\hline
\end{tabular}
\end{center}
\end{table}
As can be seen in Table 4, at the (3, -3) and (3, -1) CPs, with rVV10$d$, 
the basic ED properties are very similar to those obtained by the standard 
rVV10 functional. However, the effect of replacing rVV10 with rVV10$d$ is 
evident looking at Figure 5: one can appreciate a significant ED accumulation
along the S-S line, which is not present using the other DFT functionals
considered, including that (SCAN+rVV10) adopted by 
Kasai \textit{et al.}\cite{Kasai}
Although the discrepancy with the experimental description is only partially
resolved (even with rVV10$d$ the ED accumulation along the S-S line is 
predicted to be slightly smaller than along the S-center line), a clear
improvement is evident, which we attribute to the increased flexibility of the new S pseudopotentials that is able to better
reproduce the variety of bonding structures and hybridization processes
involving S atoms. 
\begin{table}[h!]
\caption{\scriptsize Energetic and structural properties of TiS$_2$, together with the atomic basin properties of the S atom, obtained with the rVV10$d$ and compared with rVV10. The notation used is the same as in Table 3.}
\begin{center}
\begin{adjustbox}{max width=\textwidth}
\begin{tabular}{l| c| c| c| c| c| c| c| c}
&CE (\si{\electronvolt})&ILBE (m\si{\electronvolt}/\si{\angstrom}$^2$)
&$c$ (\si{\angstrom})&CE$^{*}$ (\si{\electronvolt})&ILBE$^{*}$ (m\si{\electronvolt}/\si{\angstrom}$^2$)&V (\si{\angstrom}$^3$)&q$_{A}$ (e)&d(q$_{A}$) (e\si{\angstrom})\\
\hline
rVV10&-15.602&-27.828&5.728&-15.603&-27.888&23.775&-0.860&0.380\\
rVV10$d$&-15.640&-20.888&5.809&-15.646&-20.771&23.761&-0.860&0.390\\
\hline
Experiment&-14.805\cite{ce}& &5.665\cite{Kasai}&-14.805& &23.470\cite{Kasai}&-0.820\cite{Kasai}&0.030\cite{Kasai}\\
Theory (RPA)& &-18.900\cite{bj}& & &-18.900 &23.630\cite{Kasai}&-0.800\cite{Kasai}&0.340\cite{Kasai}\\
Theory (MBD)& &-19.031& & & -19.031& & & \\
\hline
\end{tabular}
\end{adjustbox}
\end{center}
\end{table}

Remarkably, with rVV10$d$, an even more substantial improvement in energetic properties
takes place. In fact, while the CE remains very close to that obtained by
rVV10, suggesting that $d$ orbitals of S do not play a significant role in the 
formation of strong, intralayer covalent bonds, the change in the 
ILBE is instead considerable: assuming the experimental structure, our
computed ILBE differs by only $\sim$ 2 m\si{\electronvolt}/\si{\angstrom}$^2$ from the theoretical references, i.e. high quality, RPA,\cite{bj} and MBD estimates, which represents a relative error of about 10 \%; such an error is comparable to that obtained by the much more expensive meta-GGA approaches and is much smaller than that
(about 47 \%) found with the original rVV10 functional which clearly overbinds. 
\begin{figure}[h!]
\begin{center}
\includegraphics[width=\linewidth]{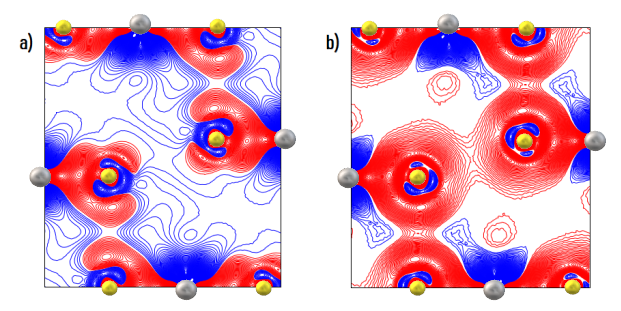}
\end{center}
\caption{ \scriptsize Comparison between the ED deformation obtained by the standard rVV10 functional (panel a) and rVV10$d$ (panel b). The smallest isoline value reported is 0.001 e/\si{\angstrom}$^3$, with red and blue contours that represent charge accumulation and depletion, respectively.}
\end{figure}
As can be seen in Table 5, the good energetic description of rVV10$d$
is preserved even considering a TiS$_2$ configuration where the experimental
lattice constant $c$ is replaced by the $c$ value optimized with rVV10$d$ (whose prediction is only slightly worse than the rVV10 one), thus increasing the confidence in the reliability of this functional.

\begin{table}[h!]
\caption{\scriptsize Optimized lattice constant $c$, ILBE and ILBE* computed for TiS$_2$ and three other TMDs, employing both rVV10 and rVV10$d$ functionals.
 The last two lines report the mean error(ME) and the mean absolute 
 relative error (MARE).}
\begin{center}
\begin{adjustbox}{max width=\textwidth}
\begin{tabular}{l| c| c| c| c| c| c}
&\multicolumn{3}{c}{rVV10$d$}&\multicolumn{3}{c}{rVV10}\\
&ILBE (m\si{\electronvolt}/\si{\angstrom}$^2$)&$c$ (\si{\angstrom})& ILBE* (m\si{\electronvolt}/\si{\angstrom}$^2$)&ILBE (m\si{\electronvolt}/\si{\angstrom}$^2$)&$c$ (\si{\angstrom})& ILBE* (m\si{\electronvolt}/\si{\angstrom}$^2$)\\
\hline
TiS$_2$&-20.771&5.809&-20.888&-27.828&5.728&-27.888\\
ref.\citenum{bj}&-18.900&5.665&-18.900&-18.900&5.665&-18.900\\
\hline
MoS$_2$&-20.296&12.810&-22.147&-29.242&12.530&-29.683\\
ref.\citenum{bj}&-20.530&12.300& -20.530& -20.530& 12.300&-20.530\\
\hline
TaS$_2$&-18.924&6.152&-19.344&-29.367&6.009&-29.369\\
ref.\citenum{bj}&-17.680&5.900& -17.680&-17.680&5.900& -17.680\\
\hline
HfS$_2$&-18.391&5.938&-18.530&-23.452&5.858&-23.603\\
ref.\citenum{bj}&-16.130&5.840& -16.130&-16.130&5.840& -16.130\\
\hline
\hline
ME&-1.285&0.251&-1.917&-9.162&0.105&-9.362\\
\hline
MARE (\%)&7.5  &3.2   &10.7  &50.3 &1.3 &51.1\\
\hline
\end{tabular}
\end{adjustbox}
\end{center}
\end{table}
As far as the dipole moment is concerned, with rVV10$d$ this quantity
is not improved with respect to the experimental estimate (see
Table 5), which can be rationalized as follows:
as a consequence of applying the modified rVV10d functional,
the electronic charge is more widespread around the S atoms,
which leads to a better agreement with experimental ED findings
and to an improved estimate of the ILBE. However this increased
charge delocalization does not occur exclusively in the interlayer
region, but also, for instance, in the intralayer region, particularly
along lines connecting neighboring S atoms. Therefore, the net
delocalization effect (see Figure 5) is approximately isotropic so that
the S dipole moment is not appreciably changed from
that estimated by the original rVV10 functional.

We have also performed additional calculations on TiS$2$ by
adopting vdW-corrected DFT functionals different from rVV10.
Interestingly, while the inclusion of the new pseudopotential involving
$d$ orbitals leads to improvements similar to those observed with
rVV10 in the description of the ED profiles
(with a small charge accumulation in the interlayer region)
and in the ILBE estimate, using vdW-corrected DFT functionals
based on non-local functionals, such as VDW-DF-cx,\cite{VDW-DF-cx}
the same is not the case when vdW-corrected DFT functionals
are adopted, which are more empirical in character (and not directly
dependent on the detailed electron density distribution), such as the
DFT-D3 approach\cite{DFT-D3}, which indicates that the new pseudopotential
is indeed better suited to describe the subtle charge deformations
occurring in this system.
Finally, to further assess the validity of this novel approach, we have applied the rVV10$d$ functional to other three different TMDs, whose chalcogen is always the S atom: MoS$_2$, TaS$_2$, and HfS$_2$. The results for the equilibrium lattice parameter $c$, as well as the ILBE and ILBE* (see Table 6 ), are very promising. The major flexibility of the new pseudopotential, embedded in rVV10$d$ scheme, dramatically improves the energetic description of S-based TMDs, maintaining an appreciable quality of the estimate of the longitudinal lattice parameter $c$: in fact, although the error relative to $c$
is $\sim$ 2 \% bigger than with rVV10, the mean absolute relative 
error (MARE) of the ILBE is reduced by about \textit{five} times; this means that
the pronounced overbinding tendency of rVV10 is almost totally eliminated. 

\section{Conclusions}
\noindent
We have presented the results of a first principles, DFT investigation, 
of the energetic and electronic properties of TiS$_2$, that is a system 
where a puzzling discrepancy, between the distribution of the electron 
density obtained by X-ray diffraction data and that computed by
state-of-the DFT schemes, has been recently reported, mainly concerning
the interlayer region between S atoms belonging to adjacent layers.
We basically confirm this observation, using the LDA, the (semilocal GGA) 
PBE, the (vdW-corrected) rVV10 DFT functionals, and also a modified rVV10 
functional (rVV10($b$=10.0)) where the $b$ parameter has been increased so
that short/medium-range vdW interactions are artificially damped.
If instead the $b$ parameter is considerably reduced to $b$=1.0 
(rVV10($b$=1.0)) significant changes occur: a stronger ED deformation in
the interlayer region is accompanied by a dramatic (and physically 
unrealistic) overestimate of the CE and ILBE terms. Substantially 
decreasing (by a 4.5 factor) the other ($C$) rVV10 parameter 
and further adjusting the $b$ parameter (rVV10($b$=1.1,$C$=0.00207)
functional) only slightly reduces the dramatic
overbinding of rVV10($b$=1.0) scheme.   
Interestingly, a more substantial improvement, in the direction of
making the theoretical description in closer agreement with the 
experiment, is obtained by adopting for S a modified, more flexible, 
pseudopotential, involving $d$ orbitals (rVV10$d$ functional).
Although the rVV10$d$ estimate of the atomic S dipole moment 
still remains substantially larger (see Table 5) than
the experimental one, a consistent improvement in the description of
the ED deformation in the interlayer region and of the ILBE is achieved.
Probably, the residual discrepancy with the experiment could be only
eliminated by an higher-level theoretical, first-principle approach,
such as a Quantum Monte Carlo scheme, in principle able to include the
whole set of correlation effects. Such an approach would be however 
extremely expensive computationally because very small ED deformations 
could only be accurately reproduced by very long simulations to get a
sufficiently small statistical error associated to differential densities.   
In any case, although all fine details in the ED deformation map cannot be
captured by rVV10$d$, nonetheless we suggest that this DFT functional 
indeed represents a reasonable compromise between accuracy and efficiency
for improving the theoretical description of TiS$_2$. Additional 
investigations indicate that similar improvements, particularly in the
evaluation of energetic terms, can be obtained by DFT calculations 
based on the rVV10$d$ functional, applied also to other TMD materials
containing S atoms and characterized by the same structure of TiS$_2$,
such as TaS$_2$, HfS$_2$, and MoS$_2$. 

\section{Supporting information}
\section{Intralayer bonding}
The analysis of the Ti-S covalent bond, which characterize the intralayer interactions, is presented here. The electron density (ED) (panel a) ) and Laplacian (panel b) ) profiles computed along the Ti-S line, as well as the ED deformation (panel c) ) and the laplacian map (panel d) ) of the Ti-S intralayer plane, are reported. Since the results achieved with all the functional schemes employed are very close to each other, we show only the PBE results as an example (Figure S1). In Table S1, the characterization of the two (3, -3) CPs and the (3, -1) CP is presented.
As can be seen, very similar values are predicted by
all the DFT functionals applied and these are also in good agreement with
experimental data, with the exception of the LDA estimate of the Laplacian of (3, -3) CP \#1, that exhibits a larger error and the Laplacian values at the (3, -1) CP, which are also substantially overestimated.
\begin{figure}[!]
\begin{center}
\includegraphics[width=14cm, height = 9.3cm]{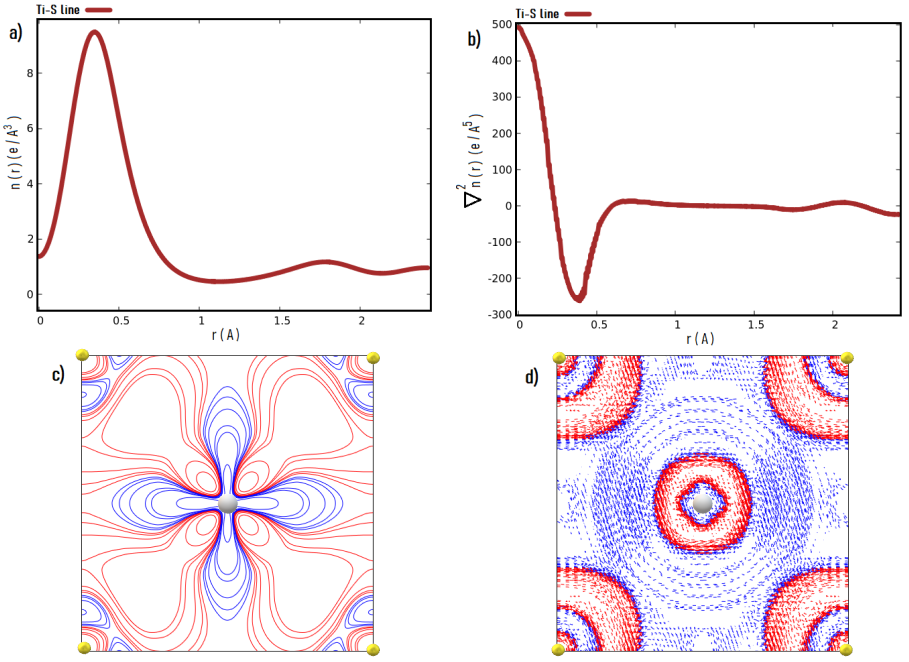}
\end{center}
\caption{\scriptsize Ti-S intralayer bond. a) ED along the Ti-S line; local maxima denote the positions of the (3,-3) CPs, while the local minimum that of the (3,-1) one; b) Laplacian of the ED; c) ED deformation density map in the Ti-S plane; d) Laplacian map in the Ti-S plane. Here results obtained at the PBE level are reported as an 
example.}
\end{figure}
\newpage

\begin{table}[!]
\caption{\scriptsize Properties at the (3, -3) CPS and (3, -1) CP identified along the Ti-S line.}
\begin{center}
\begin{adjustbox}{max width=\textwidth}
\begin{tabular}{ l| c| c| c }
(3, -3) CP \#1 &\textbf{r}$_{max}$ (\si{\angstrom}) & $n$(\textbf{r}$_{max}$) (e/\si{\angstrom}$^3$)& $\nabla^{2}n$(\textbf{r}$_{max}$) (e/\si{\angstrom}$^5$)\\
\hline
rVV10&0.353&9.490&-253.367\\
rVV10($b$=1.0)&0.350&9.487&-251.214\\
rVV10($b$=10.0)&0.353&9.495&-252.613\\
PBE&0.352&9.497&-257.543\\
LDA&0.340&9.217&-221.444\\
rVV10$opt$&0.349& 9.477&-255.304\\
\hline
Experiment\cite{Kasai}&0.407&9.350&-253.000\\
Theory\cite{Kasai} (SCAN+rVV10) &0.407&9.310&-250.000\\
\hline
\hline

(3, -3) CP \#2 &\textbf{r}$_{max}$ (\si{\angstrom}) & $n$(\textbf{r}$_{max}$) (e/\si{\angstrom}$^3$)& $\nabla^{2}n$(\textbf{r}$_{max}$) (e/\si{\angstrom}$^5$)\\
\hline
rVV10&1.777&1.171&-10.315\\
rVV10($b$=1.0)&1.782&1.175&-9.978\\
rVV10($b$=10.0)&1.780&1.175&-10.789\\
PBE&1.781&1.172&-10.430\\
LDA&1.819&1.216&-10.091\\
rVV10$opt$&1.778&1.157&-10.165\\
\hline
Experiment\cite{Kasai}&1.714&1.260&-11.300\\
Theory\cite{Kasai}(SCAN+rVV10)&1.707&1.200&-9.870\\
\hline
\hline
(3, -1) CP &\textbf{r}$_{min}$ (\si{\angstrom}) & $n$(\textbf{r}$_{min}$) (e/\si{\angstrom}$^3$)& $\nabla^{2}n$(\textbf{r}$_{min}$) (e/\si{\angstrom}$^5$)\\
\hline
rVV10&1.122&0.453&1.455\\
rVV10($b$=1.0)&1.126&0.448&1.211\\
rVV10($b$=10.0)&1.122&0.450&1.484\\
PBE&1.120&0.458&1.231\\
LDA&1.125&0.461&1.002\\
rVV10$opt$&1.124&0.453&1.245\\
\hline
Experiment\cite{Kasai}& &0.429&3.791\\
Theory\cite{Kasai} (SCAN+rVV10)& &0.421&3.956\\
\hline
\end{tabular}
\end{adjustbox}
\end{center}
\end{table}

\newpage
\section{ED interlayer Laplacians}
\noindent
Here we present the Laplacian maps of the interlayer plane of interest, computed for all the functional employed (the LDA map is very similar to the ones shown).
\begin{figure}[!]
\begin{center}
\includegraphics[width=\linewidth]{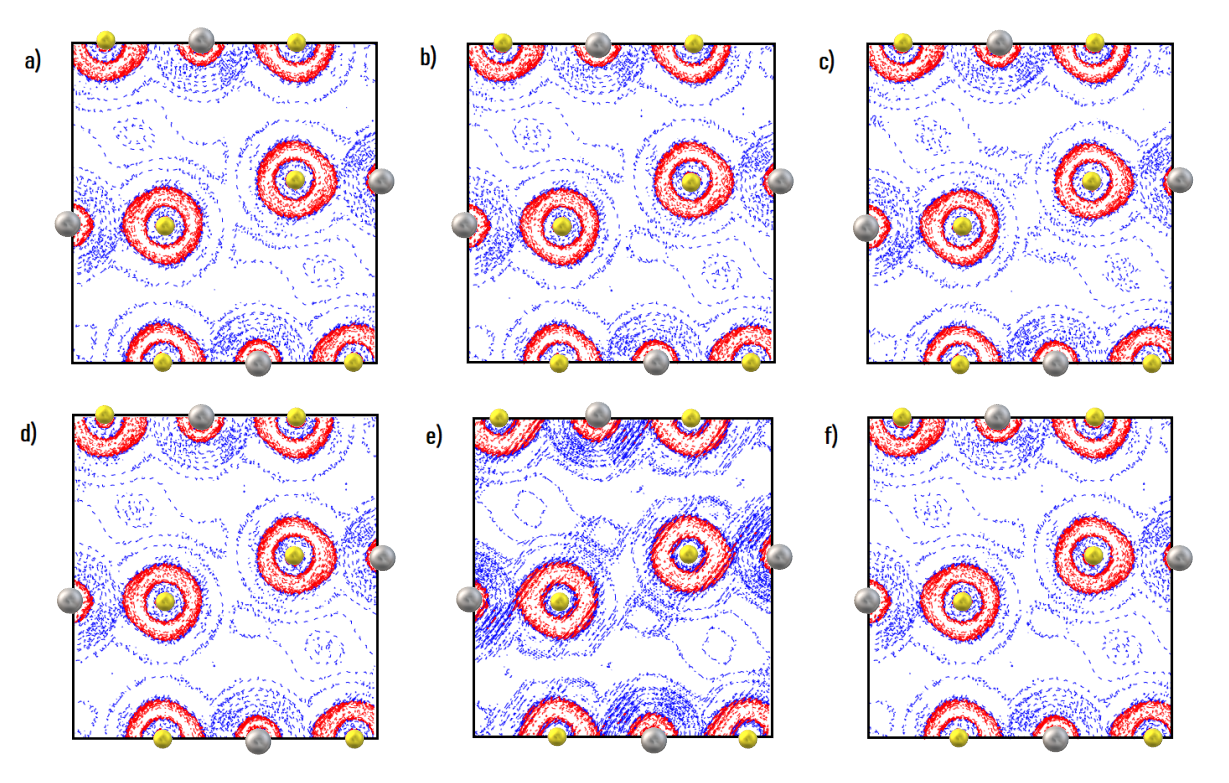}
\end{center}
\caption{\scriptsize Laplacian map in the interlayer plane using different functionals: a) PBE, b) rVV10, c) rVV10($b$=1.0), d) rVV10($b$=10.0), e) rVV10$opt$, and f) rVV10$d$. The Ti atom are represented by grey spheres, while the S atoms are the yellow spheres. The results are reported by using Bader contour lines of [1, 2, 4, 8]$\times$ 10 $^{[-3, \,-2, \,-1, \,0, \,1]}$ e/\si{\angstrom}$^5$. In these maps red (blue) contours are associated to negative (positive) laplacian values, which represents charge accumulation (depletion).}
\end{figure}
\newpage
\section{S$_2$ molecule characterization}
\noindent
The calculations for the S$_2$ molecule, were carried out using the same 
approach adopted for the simulations of the TiS$_2$ system (see 
section 2 in the main text); the only differences were the chosen 
supercell (a cube with sides of 13.2 \si{\angstrom}) and the sampling
of the BZ limited to the $\Gamma$ point. 
Calculations were spin-polarized and the 
S$_2$ vibrational frequency was computed using the \textit{PHonon} program of QE.
\begin{table}[!]
\caption{\scriptsize Equilibrium distance ($d$), binding energy, and vibrational frequency of the S$_2$ molecule, computed with the rVV10 and rVV10$d$ functionals.}
\begin{center}
\begin{tabular}{ l| c| c| c}
& $d$ (\si{\angstrom}) & binding energy (\si{\electronvolt})& vibrational frequency (cm$^{-1}$)\\
\hline
rVV10&1.91&5.73&689\\
rVV10$d$&1.91&4.57&692\\
\hline
Experiment&1.89\cite{dstate}&4.74\cite{dstate}& 714 $\pm$ 12\cite{s2}\\
\hline
\end{tabular}
\end{center}
\end{table}

\section{Acknowledgements}
We acknowledge funding from Fondazione Cariparo, Progetti di Eccellenza 2017,
relative to the project: ''Engineering van der Waals
Interactions: Innovative paradigm for the control of Nanoscale
Phenomena''.

\vfill
\eject

\begin{figure}
\label{TOC Graphic}
\begin{center}
\includegraphics[width=\linewidth]{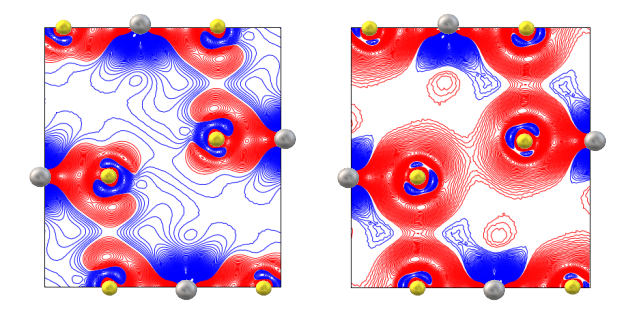}
\end{center}
\end{figure}
\vfill
\eject


This is an auxiliary file used by the `achemso' bundle.
This file may safely be deleted. It will be recreated as required.
 
@Control{achemso-control,
  ctrl-article-title  = "yes",
  ctrl-chapter-title  = "no",
  ctrl-doi            = "no",
  ctrl-etal-number    = "0",
  ctrl-etal-firstonly = "yes",
}



\newpage
\begin{thebibliography}{9}
\bibitem{vdW} Van der Waals, J.D. Over de Continuiteit van den Gas- en Vloeistoftoestand (on the continuity of the gas and liquid state). \textit{PhD thesis} \textbf{1873}.
\bibitem{LM} Duong, D. L.; Yun, S. J.; Lee, Y. H. van der Waals Layered Materials: Opportunities and Challenges. \textit{ACS Nano} \textbf{2017}, \textit{11}, 11803-11830.
\bibitem{LDA} Kohn, W.; Sham, L. J. Self-Consistent Equations Including Exchange and Correlation Effects. \textit{Phys. Rev.} \textbf{1965}, \textit{140}, A1133-A1138.
\bibitem{PBE} Perdew, J. P.; Burke, K.; Ernzerhof, M. Generalized Gradient Approximation Made Simple. \textit{Phys. Rev. Lett.} \textbf{1996}, \textit{77}, 3865-3868; 
Erratum: \textit{Phys. Rev. Lett.} \textbf{1997}, \textit{78}, 1396.
\bibitem{Kohn} Kohn, W.; Meir, Y.; Makarov, D. E.
van der Waals Energies in Density Functional Theory.
{\it Phys. Rev. Lett.} \textbf{1998}, {\it 80}, 4153-4156.
\bibitem{Pernal} Pernal, K.; Podeszwa, R.; Patkowski, K.; Szalewicz, K.
Dispersionless Density Functional Theory.
{\it Phys. Rev. Lett.} \textbf{2009}, {\it 103}, 263201.
\bibitem{Stohr} St\"ohr, M.; Van Voorhis, T.; Tkatchenko, A.
Theory and practice of modeling van der Waals interactions in
electronic-structure calculations.
{\it Chem. Soc. Rev.} \textbf{2009}, {\it 48}, 4118-4154.
\bibitem{Ruiz} Ruiz, V.G.; Liu, W.; Zojer, E.; Scheffler, M.; Tkatchenko, A.
Density-functional theory with screened van der Waals interactions for the
modeling of hybrid inorganic-organic systems.
{\it Phys. Rev. Lett.} \textbf{2012}, {\it 108}, 146103.
\bibitem{Su} Su, G.; Yang, S.; Li, S.; Butch, C. J.; Filimonov, S. N.; Ren, J.-C.;
Liu, W.
Modeling chemical reactions on surfaces: The roles of chemical bonding
and van der Waals interactions.
{\it Prog. Surf. Sci.} \textbf{2019}, {\it 94}, 100561. 
\bibitem{Foulkes} Foulkes, W.; Mitas, L.; Needs, R.; Rajagopal, G. Quantum Monte Carlo Simulations of Solids. \textit{Rev. Mod. Phys.} \textbf{2001}, \textit{73}, 33-83.
\bibitem{Raghavachari} Raghavachari, K.; Trucks, G. W.; Pople, J. A.; Head-Gordon, M. A Fifth-Order Perturbation Comparison of Electron Correlation Theories. \textit{Chem. Phys. Lett.} \textbf{1989}, \textit{157}, 479-483.
\bibitem{Eshuis} Eshuis, H.; Bates, J. E.; Furche, F. Electron Correlation Methods Based on the Random Phase Approximation. \textit{Theor. Chem. Acc.} \textbf{2012}, \textit{131}, 
1089-1102.
\bibitem{Medvedev} Medvedev, M. G.; Bushmarinov, I. S.; Sun, J.; Perdew, J. P.; Lyssenko, K. A. Density functional theory is straying from the path toward the exact functional. \textit{Science} \textbf{2017}, \textit{355}, 49-52.
\bibitem{Kasai} Kasai, H.; Tolborg, K.; Sist M.; Zhang, J.; Hathwar, V. R.; Filsø, M. Ø.; Cenedese, S.; Sugimoto, K.; Overgaard J.; Nishibori E. et al. X-ray electron density investigation of chemical bonding in van der Waals materials. \textit{Nature Materials} \textbf{2018}, \textit{17}, 249-253.
\bibitem{rVV10} Sabatini R.; Gorni, T.; de Gironcoli, S. Nonlocal van der Waals density functional made simple and efficient. \textit{Phys. Rev. B} \textbf{2013}, \textit{87}, 041108.
\bibitem{PAW} Blöchl, P. E. Projector augmented-wave method. \textit{Phys. Rev. B} \textbf{1994}, \textit{50}, 17953-17979.
\bibitem{dstate} Yourdshahyan, Y.; Zhang, H. K.; Rappe, A. M. n-alkyl thiol head-group interactions with the Au(111) surface. \textit{Phys. Rev. B} \textbf{2001}, \textit{63}, 081405.
\bibitem{grimme1} Grimme, S. Accurate description of van der Waals complexes by density functional theory including empirical corrections. \textit{J. Comput. Chem.} \textbf{2004}, \textit{25}, 1463-1473.
\bibitem{grimme2} Grimme, S. Semiempirical GGA‐type density functional constructed with a long‐range dispersion correction. \textit{J. Comput. Chem.} \textbf{2006}, \textit{27}, 1787-1799.
\bibitem{xdm} Becke, A. D.; Johnson, E. R. Exchange-hole dipole moment and the dispersion interaction: High-order dispersion coefficients. \textit{J. Chem. Phys.} \textbf{2006}, \textit{124}, 014104.
\bibitem{Dion} Dion, M.; Rydberg, H.; Schröder, E.; Langreth, D. C.; Lundqvist, B. I. Van der Waals Density Functional for General Geometries. \textit{Phys. Rev. Lett.} \textbf{2004}, \textit{92}, 246401.
\bibitem{VV10} Vydrov, O. A.; Van Voorhis, T. Nonlocal van der Waals density functional: The simpler the better. \textit{J. Chem. Phys.} \textbf{2010}, \textit{133}, 244103.
\bibitem{rPW86} Murray, É. D.; Lee, K.; Langreth, D. C. Investigation of Exchange Energy Density Functional Accuracy for Interacting Molecules. \textit{J. Chem. Theory Comput.} \textbf{2009}, \textit{5}, 2754-2762.
\bibitem{LDAc} Perdew, J. P.; Wang, Y. Accurate and simple analytic representation of the electron-gas correlation energy. \textit{Phys. Rev. B} \textbf{1992}, \textit{45}, 
13244-13249.
\bibitem{BE} Jurečka, P.; Šponer, J.; Černý, J.; Hobza, P. Benchmark database of accurate (MP2 and CCSD(T) complete basis set limit) interaction energies of small model complexes, DNA base pairs, and amino acid pairs. \textit{Phys. Chem. Chem. Phys.} \textbf{2006}, \textit{8}, 1985-1993.
\bibitem{RPS} Román-Pérez, G.; Soler, J. M. Efficient Implementation of a van der Waals Density Functional: Application to Double-Wall Carbon Nanotubes. \textit{Phys. Rev. Lett.} \textbf{2009}, \textit{103}, 096102.
\bibitem{QE1} Giannozzi, P.; Baroni, S.; Bonini, N.; Calandra, M.; Car, R.; Cavazzoni, C.;  Ceresoli, D.; Chiarotti, G. L.; Cococcioni, M.; Dabo, I. QUANTUM ESPRESSO: a modular and open-source software project for quantum simulations of materials. \textit{Journal of Physics: Condensed Matter} \textbf{2009}, \textit{21}, 395502-395521.
\bibitem{QE2} Giannozzi, P.; Andreussi, O.; Brumme, T.; Bunau, O.; Buongiorno Nardelli, M.; Calandra, M.; Car, R.; Cavazzoni, C.; Ceresoli, D.; Cococcioni, M. et al. Advanced capabilities for materials modelling with Quantum ESPRESSO. \textit{Journal of Physics: Condensed Matter} \textbf{2017}, \textit{29}, 465901-465931 .
\bibitem{pseudo} http://www.quantum-espresso.org/pseudopotentials
\bibitem{spline} Charles, H. A.; Meyer, W. W. Optimal Error Bounds for Cubic Spline Interpolation. \textit{Journal of Approximation Theory} \textbf{1976}, \textit{16}, 
105-122.
\bibitem{Critic2_1} Otero-de-la-Roza A.; Blanco, M. A.; Pendás, A. M.; Luaña, V. Critic: a new program for the topological analysis of solid-state electron densities. \textit{Comput. Phys. Commun.} \textbf{2009}, \textit{180}, 157-166. 
\bibitem{Critic2_2} Otero-de-la-Roza, A.; Johnson, E. R.; Luaña, V. Critic2: a program for
real-space analysis of quantum chemical interactions in solids. \textit{Comput. Phys. Commun.} \textbf{2014}, \textit{185}, 1007-1018 .
\bibitem{SCAN+rVV10} Peng, H.; Yang, Z. H.; Perdew, J. P.; Sun, J. Versatile van der
Waals Density Functional Based on a Meta-Generalized Gradient Approximation. \textit{Phys. Rev. X} \textbf{2016}, \textit{6}, 041005.
\bibitem{bj} Björkman, T.; Gulans, A.; Krasheninnikov, A. V.; Nieminen, R. M. van der Waals Bonding in Layered Compounds from Advanced Density-Functional First-Principles Calculations. \textit{Phys. Rev Lett.} \textbf{2012}, \textit{108}, 235502.
\bibitem{ce} Clerc, D. G.; Poshusta, R. D.; Hess, A. C. Periodic Hartree-Fock Study of TiS$_2$. \textit{J. Phys. Chem.} \textbf{1996}, \textit{100}, 15735-15747.
\bibitem{mbd} Ambrosetti, A.; Reilly, A. M.; DiStasio Jr., R. A.; Tkatchenko, A.
Long-range correlation energy calculated from coupled atomic response
functions.
{\it J. Chem. Phys.} \textbf{2014}, {\it 140}, 18A508.
\bibitem{prb} Ambrosetti, A.; Silvestrelli, P.L.; Tkatchenko, A.
Physical adsorption at the nanoscale: Towards controllable scaling of
the substrate-adsorbate van der Waals interaction.
{\it Phys. Rev. B} \textbf{2017}, {\it 95}, 235417. 
\bibitem{LDAprob} Becke, A. D. Perspective: Fifty years of density-functional theory in chemical physics. \textit{J. Chem. Phys.} \textbf{2014}, \textit{140}, 
18A301.
\bibitem{tm} Troullier, N.; Martins, J. L. Efficient pseudopotentials for plane-wave calculations. \textit{Phys. Rev. B} \textbf{1991}, \textit{43}, 1993-2006.
\bibitem{nlcc} Louie, S. G.; Froyen, S.; Cohen, M. L. Nonlinear ionic pseudopotentials in spin-density-functional calculations. \textit{Phys. Rev. B} \textbf{1982}, \textit{26}, 1738-1742.
\bibitem{s2} Qin, Z.; Wang, L.; Cong, R.; Jiao, C.; Zheng, X.; Cui, Z.;
Tang, Z. Spectroscopic identification of the low-lying states of S$_2$ molecule. 
\textit{J. Chem. Phys.} \textbf{2019}, \textit{150}, 044302.
\bibitem{VDW-DF-cx} Berland, K.; Hyldgaard, P.
Exchange functional that tests the robustness of the plasmon
description of the van der Waals density functional.
{\it Phys. Rev. B} \textbf{2014}, {\it 89}, 035412.
\bibitem{DFT-D3} Grimme, S.; Antony, J.; Ehrlich, S.; Krieg, H.
A consistent and accurate ab initio parametrization of density
functional dispersion correction (DFT-D) for the 94
elements H-Pu.
{\it J. Chem. Phys.} \textbf{2010}, {\it 132}, 154104.
\end{thebibliography}
\end{document}